\documentclass[onecollarge]{svjour2}       
\smartqed  
\usepackage{graphicx}
\begin{document}

\title{Tendency to occupy a statistically dominant spatial state
of the flow as a driving force for turbulent transition}%

\titlerunning{Driving force for turbulent transition}        

\author{Sergei F. Chekmarev}


\institute{Sergei F. Chekmarev \at Institute of Thermophysics, SB RAS, 630090 Novosibirsk,
Russia, and\\ Department of Physics, Novosibirsk State University, 630090 Novosibirsk, Russia \\
              Tel.: +7(383)3165048\\
              Fax: +7(383)3308480\\
              \email{chekmarev@itp.nsc.ru}}

\date{Received: date / Accepted: date}

\maketitle

\begin{abstract}
A simple analytical model for a turbulent flow is proposed, which considers the flow as
a collection of localized spatial structures that are composed of elementary "cells"
in which the state of the particles (atoms or molecules) is uncertain. The Reynolds
number is associated with the ratio between the total phase volume for the system
and that for the elementary cell. Calculating the statistical weights of the collections
of the localized structures, it is shown that as the Reynolds number increases, the
elementary cells group into the localized structures, which successfully explains the
onset of turbulence and some other characteristic properties of turbulent flows. It is
also shown that the basic assumptions underlying the model are involved in the derivation
of the Navier-Stokes equation, which suggests that the driving force for the turbulent
transition described with the hydrodynamic equations is essentially the same as in the
present model, i.e. the tendency of the system to occupy a statistically dominant state
plays a key role. The instability of the flow can then be a mechanism to initiate the
structural rearrangement
of the flow to find this state.%

\end{abstract}

\section{Introduction}
\label{intro} The transition from laminar to turbulent fluid motion occurring at large
Reynolds numbers
\cite{Reynolds1883,Richardson1922,MoninYaglom,LandauLifshitz_fluid_mech,Lesieur} is
generally associated with the instability of the laminar flow, and this viewpoint is well
supported by the analysis of solutions of hydrodynamic equations
\cite{LandauLifshitz_fluid_mech,Lesieur,Drazin02}. On the other hand, since the turbulent
flow characteristically appears in the form of eddies filling the flow field, the
tendency to occupy such a structured state of the flow cannot be ruled out as a driving
force for turbulent transition. To examine this possibility, we propose a simple
analytical model for the flow and show that as the Reynolds number increases, the state
of the flow in the form of collection of localized spatial structures (eddies) becomes
statistically more favorable than the unstructured state, which successfully explains the
onset of turbulence and some other general properties of turbulent flows. We also show
that the basic assumptions underlying the model are involved in the derivation of
the Navier-Stokes equation, which suggests that the driving force for the turbulent
transition is basically the same as in the present model, i.e. the tendency of the system
to occupy a statistically dominant state plays a key role. The instability of the flow at
high Reynolds numbers can then be a mechanism to initiate the structural rearrangement of
the flow to find this state.

\section{Model}
\label{sec:1}%
Let a fluid flow be represented by a system of $N$ identical particles (atoms or
molecules). Assume that the particles can form localized structures of $N_i$ particles
(with the number of such structures being $M_i$), which are similar to turbulent eddies
in that the particles execute an overall concerted motion. Since the concerted motion
should break down at the microscale level, each $i$ structure can be divided into
$N_{i}/n$ elementary "cells" (each of $n$ particles), in which the states of the
particles (positions and velocities) are uncorrelated. Then, considering the
interaction between the localized structures to be small and taking into account that the
permutations of the particles in the elementary cells, those of the elementary cells in
the localized structures, and the permutations of the localized structures themselves do
not lead to new states, the total number of states can be estimated as for an ideal-gas
system \cite{LandauLifshitz_stat_phys}
$$\Gamma=\frac{N!}{\prod_{i}\left[(n!)^{N_{i}/n}(N_{i}/n)!\right]^{M_i}M_{i}!}$$
or, with the Stirling approximation $x!\approx (x/e)^x$ to be applicable
\begin{equation}\label{eq1}
 \Gamma=\frac{(N/n)^{N}e^{N/n}}{\prod_{i}(N_{i}/n)^{N_{i}M_{i}/n}(M_{i}/e)^{M_{i}}}
\end{equation}
where $\sum_{i}N_{i}M_{i}=N$.

With Eq. (\ref{eq1}), it is possible to calculate the most probable size distribution of
the structures, $\tilde{M}_{i}=\tilde{M}_{i}(N_{i})$, which maximizes the number of
states $\Gamma$ and, correspondingly, the entropy $S=\ln \Gamma$. This distribution
reduces the phase space of the system to a subspace of much lower dimension, similar to
how, according to the dynamical system theory, the phase space shrinks when the system
reaches the (strange) attractor in the transition to turbulence
\cite{LandauLifshitz_fluid_mech,RuelleTakens71}. Applying the
Lagrange multiplier method to the entropy functional, i.e. varying the expression
$\sum_{i}\left[-(M_{i}N_{i}/n)\ln (N_{i}/n)-M_{i}\ln(M_{i}/e)\right]
+\alpha'\sum_{i}M_{i}N_{i}+\beta'\sum_{i}M_{i}E_{i}$ with respect to $M_{i}$  at two
conservation conditions $\sum_{i}M_{i}N_{i}=N$ and $\sum_{i}M_{i}E_{i}=E$ ($E_{i}$ is the
kinetic energy of fluctuations in $i$ structure, and $E$ is the total kinetic energy),
one obtains
\begin{equation}\label{eq2}
\tilde{M}_{i}=\left(\frac{N_{i}}{n}\right)^{-\frac{N_{i}}{n}}e^{\alpha' N_{i}+\beta'
E_{i}}
\end{equation}
where and $\alpha'$ and $\beta'$ are the Lagrange multipliers.

The dependence of $E_{i}$ on $N_{i}$ is specific rather than universal; it is clearly
different, e.g., for homogeneous and shear-layer turbulence. To be definite, assume that
the velocity of fluctuations increases with the distance $r$ as in the Kolmogorov theory
of turbulence for the inertial range of scales
\cite{LandauLifshitz_fluid_mech,Kolmogorov41a,Frisch}, i.e.
\begin{equation}\label{eq3}
v(r) \sim r^{h}
\end{equation}
where $h=1/3$. Then, the kinetic energy of fluctuations per unit mass is $e(r) \sim
r^{2h}$, and $E_{i} \sim \int_{0}^{r_i} e(r)r^{2}dr \sim r_{i}^{2h+3} \sim
N_{i}^{2h/3+1}$. Also, it is convenient to pass from $N_{i}$ to the number of the
elementary cells in the localized structure $q_{i}=N_{i}/n$. With these changes, Eq.
(\ref{eq2}) becomes
\begin{equation}\label{eq4}
\tilde{M}_{i}=q_{i}^{-q_{i}}e^{\alpha q_{i}+\beta q_{i}^{2h/3+1}}
\end{equation}
where $\alpha$ and $\beta$ are new constants that are determined from the equations of
conservation of the total number of particles and kinetic energy
\begin{equation}\label{eq5}
N/n=\sum_{i}\tilde{M}_{i}q_{i}
\end{equation}
and
\begin{equation}\label{eq6}
E/n^{2h/3+1}=\sum_{i}\tilde{M}_{i}q_{i}^{2h/3+1}
\end{equation}

Given the values of $n$, $N$ and $E$, the constants $\alpha$ and $\beta$ can be
calculated by generating a set of $q_{i}$ and solving Eqs. (\ref{eq5}) and (\ref{eq6})
with $\tilde{M}_{i}$ substituted from Eq. (\ref{eq4}). Another possibility is to vary
$\alpha$ and $\beta$ to obtain $N/n$ and $E/n^{2h/3+1}$ as functions of $\alpha$ and
$\beta$.

The parameter $N/n$ can be associated with the Reynolds number ${\mathrm
{Re}}_{L}=VL/\nu$, where $V$ and $L$ are, respectively, the velocity and linear scale
characterizing the system as a whole, and $\nu$ is the kinematic viscosity of the fluid.
Since the particles are identical, the $6N$-dimensional phase space is reduced to the
$6$-dimensional single-particle space. Then, the linear size of the elementary cell, in
which the behavior of the particles is uncorrelated, can be taken as the product of the
characteristic velocity and length at which the state of the particles is uncertain. For
a gas fluid, it is the product of the molecular thermal velocity $c$ and the mean free
path $\lambda$ \cite{Kusukawa51}. Then, the phase volume of the cell is $(c \lambda)^3$,
with the number of particles in the cell being proportional to this volume, $n \sim (c
\lambda)^3$. Correspondingly, the total number of particles $N$ can be considered to be
proportional to the total phase volume for the system $(VL)^3$, i.e. $N \sim (VL)^3$ (the
density of the fluid is assumed to be constant). Taking into account the kinetic theory
expression for the gas viscosity $\nu \sim c \lambda$ (e.g., Ref. \cite{ChapmanCowling}),
one obtains
\begin{equation}\label{eq7}
N/n \sim (VL/\nu)^3 ={\mathrm {Re}}_{L}^3
\end{equation}
A similar relation is valid for a liquid fluid. Here, the linear size of the elementary
cell can be taken to be $c^2 \tau$, where $c^2$ presents the fluctuations of the
(molecular) kinetic energy per unit mass, and $\tau$ is the mean residence time of the
molecule in a settled state. Since the viscosity of liquid is $\nu \sim c^2 \tau$ (e.g.,
Ref. \cite{Frenkel}), we arrive at Eq. (\ref{eq7}) again. In what follows, we will assume
for simplicity that the coefficient of proportionality in Eq. (\ref{eq7}) is equal to 1,
i.e.
\begin{equation}\label{eq7a}
{\mathrm {Re}}_{L}=(N/n)^{1/3}
\end{equation}

\section{Results and Discussion}
\label{sec:2}%

\subsection{Turbulent Transition Region}
\label{sec:2.1}%
\begin{figure}
\centering
\resizebox{0.5\columnwidth}{!}{ \includegraphics* {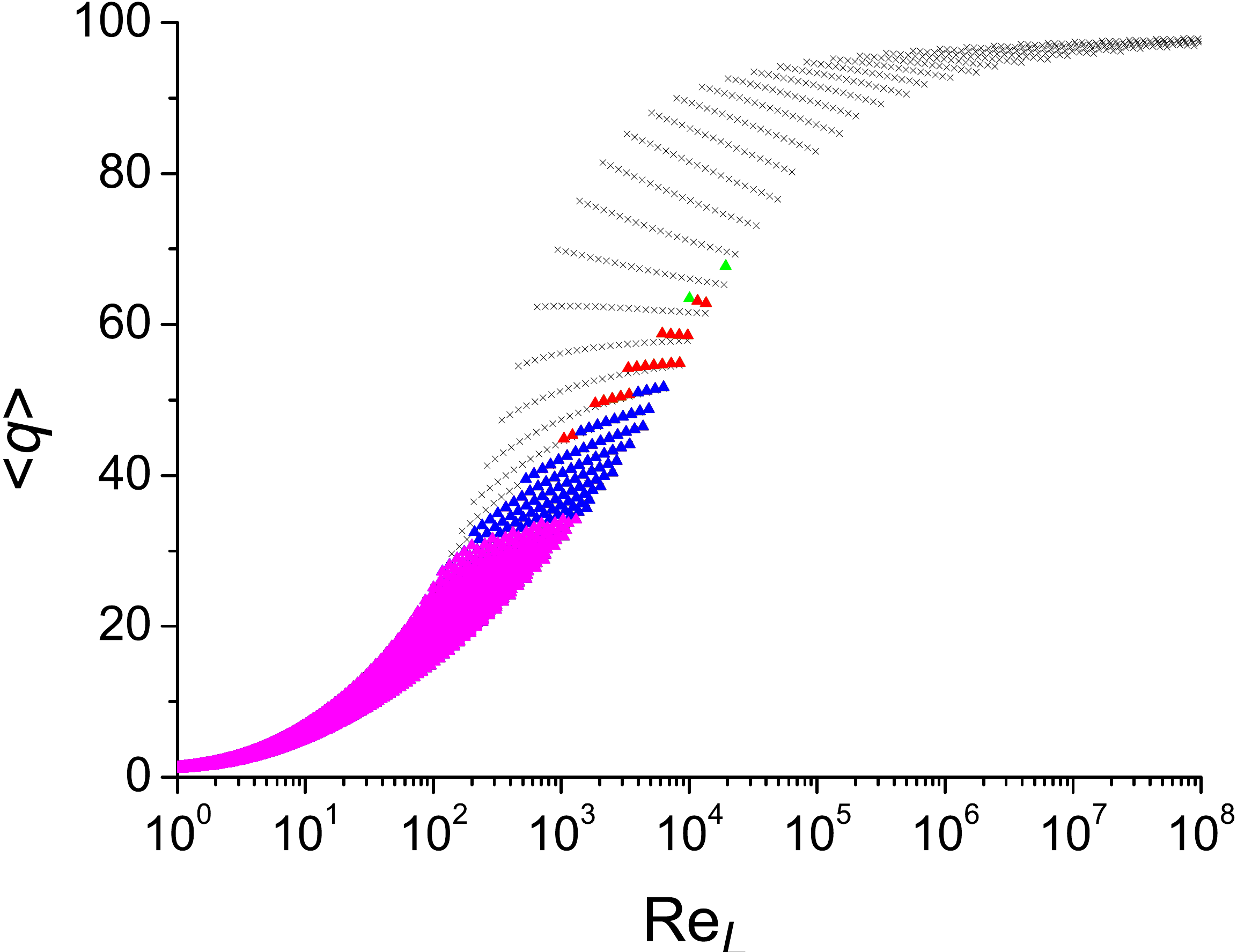}}%
\caption{(Color online) Average number of the elementary cells in the localized
structures $\langle q\rangle$ as a function of $\mathrm{Re}_{L}$. The number of the
elementary cells $q_i$ varies from 1 to $q_{\mathrm{max}}$, and parameters $\alpha$ and
$\beta$ vary independently from $\alpha_{\mathrm{min}}=\beta_{\mathrm{min}}=-1.5$ to
$\alpha_{\mathrm{max}}=\beta_{\mathrm{max}}=1.5$ with the interval 0.015. Crosses
represent all combinations of $\alpha$ and $\beta$ at $q_{\mathrm{max}}=100$, and the
triangles stand for the combinations which lead to the physically reasonable bell-shaped
distribution at different $q_{\mathrm{max}}$; the magenta, blue, red and green triangles
are for $q_{\mathrm{max}}=100, 200, 300$ and 400, respectively.}
\label{fig1}       
\end{figure}

Figure \ref{fig1} shows a characteristic dependence of the average number of elementary
cells in the localized structures $\langle
q\rangle=\sum_{i}q_{i}\tilde{M}_{i}/\sum_{i}\tilde{M}_{i}$ on the Reynolds number
${\mathrm {Re}}_{L}$; the number of elementary cells $q_i$ varies from 1 to
$q_{\mathrm{max}}$. Points labelled with crosses represent all combinations of $\alpha$
and $\beta$, with these parameters varied independently within $\alpha_{\mathrm{min}}
\leq \alpha \leq \alpha_{\mathrm{max}}$ and $\beta_{\mathrm{min}} \leq \beta \leq
\beta_{\mathrm{max}}$. However, not every combination leads to a physically reasonable
bell-shaped distribution \cite{JimenezWraySaffmanRogallo93} (see also Fig. \ref{fig4}
below). In small fraction of the
combinations (typically at large values of $\beta$), $\tilde{M}_{i}$ does
not vanish at $q \rightarrow q_{\mathrm{max}}$. The fraction of such "wrong" combinations
is generally within several percentages (in the given case, below 2 percent). Triangles
show the dependence $\langle q\rangle$ on ${\mathrm {Re}}_{L}$ with those "wrong" $\alpha
/ \beta$ combinations excluded. The increase of $q_{\mathrm{max}}$ does not change the
distribution, except that the range of variation of $\langle q\rangle$ and ${\mathrm
{Re}}_{L}$ is extended to larger values of these quantities.

It is seen that as ${\mathrm {Re}}_{L}$ exceeds some characteristic value (${\mathrm
{Re}}_{L} \sim 10^2$ in Fig. \ref{fig1}), $\langle q\rangle$ rapidly increases, i.e.
the elementary cells group into localized structures, signaling that the flow becomes
turbulent at these Reynolds numbers. It is noteworthy that the dependence
of $\langle q \rangle$ on ${\mathrm {Re}}_{L}$ is not unique, i.e. the same values of
$\langle q\rangle$ are observed in a broad range of ${\mathrm
{Re}}_{L}$. This indicates that the
\begin{figure}
\centering
\resizebox{0.5\columnwidth}{!}{ \includegraphics* {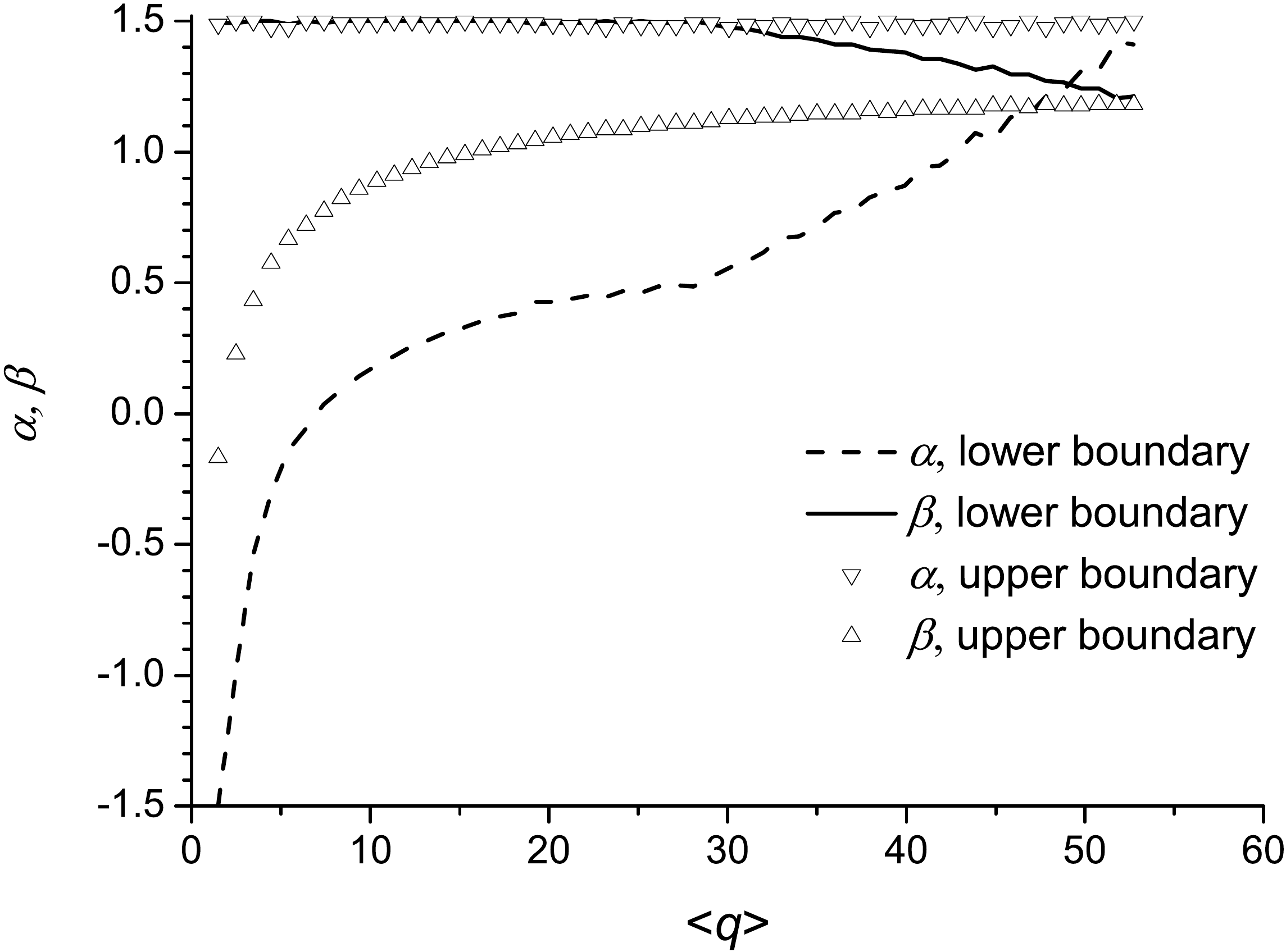}}%
\caption{The values of the parameters $\alpha$ and $\beta$ at the lower and upper
boundaries of the transition region of Fig. \ref{fig1}; $q_{\mathrm{max}}=200$.}
\label{fig2}       
\end{figure}
turbulent state is not solely determined by the Reynolds number; rather, as is
well-known, it is flow specific, depending on the type of the flow, the inlet conditions,
the flow environment, etc. \cite{MoninYaglom,LandauLifshitz_fluid_mech,Lesieur,Drazin02}.
The present statistical model is too simple to take these effects into account, but it
offers an estimate for the range of ${\mathrm {Re}}_{L}$ at which the turbulent motion
can be expected, i.e. the lower and upper bounds of the turbulent transition region. The
lower and upper boundaries of the manifold of the "correct" points in Fig. \ref{fig1}
(labelled with triangles) are determined, respectively, by the maximum values of $\beta$
and $\alpha$ at which the correct points are obtained (Fig. \ref{fig2}). As these
parameters increase, the lower boundary shifts to smaller values of ${\mathrm {Re}}_{L}$,
and the upper boundary to larger values of ${\mathrm {Re}}_{L}$ (Supplementary Material).
More specifically, at the lower boundary
${\mathrm {Re}}_{L} \sim \exp(-0.15 \beta\langle q\rangle)$, and at the upper boundary
${\mathrm {Re}}_{L} \sim \exp(0.06 \alpha\langle q\rangle)$. Figure \ref{fig3} shows
these dependencies for $\langle q\rangle=40$, which corresponds
to approximately the maxima of the size distributions of
the localized structures (see Fig. \ref{fig4} and its discussion below). It is remarkable
that the lower boundary moves until ${\mathrm {Re}}_{L} \sim 10^2$ is reached; after that
it "freezes" (Supplementary Material). The model thus predicts that the flow should be
laminar at ${\mathrm {Re}}_{L} < {\mathrm {Re}}_{L}^{\star}$ (${\mathrm {Re}}_{L}^{\star}
\sim 10^2$ in the present model), and it can remain laminar up to very large values of
${\mathrm {Re}}_{L}$. This prediction is in good general agreement with the experimental
and simulation results \cite{MoninYaglom,LandauLifshitz_fluid_mech,Lesieur}. In
particular, for the pipe flow, in his seminal work Reynolds \cite{Reynolds1883} has found
that the critical Re number varied from $\mathrm{Re}_{L} \approx 2 \times 10^3$ to
$\mathrm{Re}_{L} \approx 1.3 \times 10^4$ depending on the inlet conditions. The lower
bound of the stability of the laminar flow, which is observed at large disturbances of
the inlet flow, has been confirmed in many experimental and theoretical works (e.g.,
Refs. \cite{DarbyshireMullin95} and \cite{Ben-DavCohen07}, respectively), and the upper
bound, which is achieved in carefully controlled conditions, has been found as high as
$\mathrm{Re}_{L} \approx 1 \times 10^5$ \cite{Pfenninger61}. In the present model, the
above mentioned lower bound of the stability ($\mathrm{Re}_{L} \sim 10^3$) corresponds to
$\beta_{\mathrm{max}} \approx 1.25$, and the upper bound is unlimited (Fig. 3).
\begin{figure}
\centering
\resizebox{0.5\columnwidth}{!}{ \includegraphics* {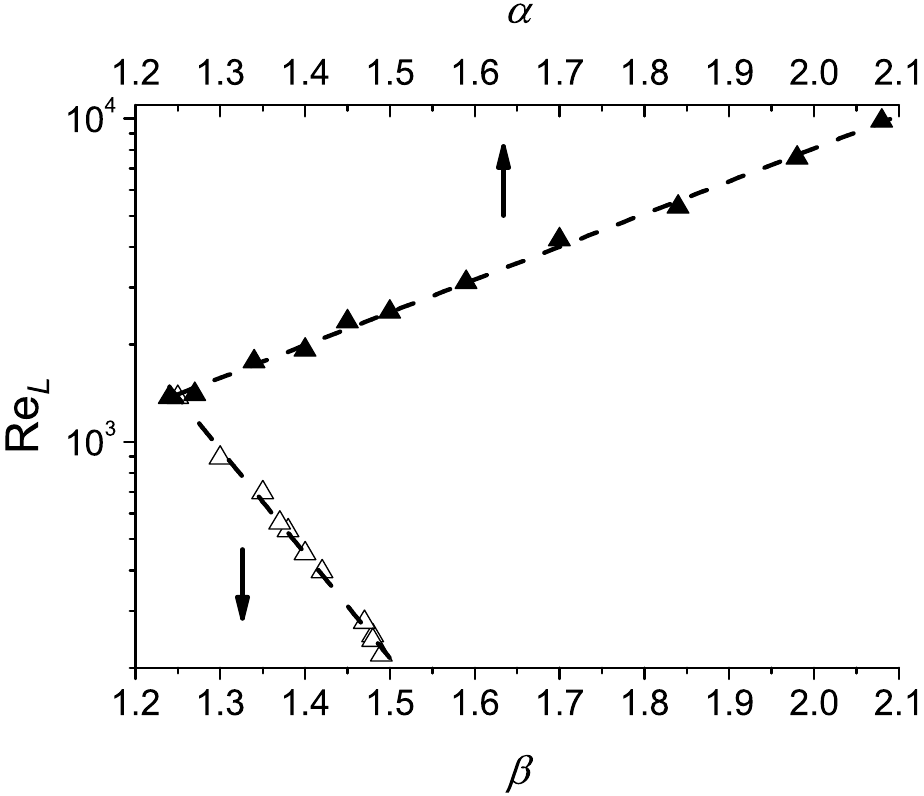}}%
\caption{The Reynolds number ${\mathrm {Re}}_{L}$ corresponding to the lower (open
triangles) and the upper (solid triangles) boundaries of the transition range as a
function of parameters $\beta$ and $\alpha$, respectively; $\langle q\rangle=40$. The
dashed lines are the best exponential fits, see the text.}
\label{fig3}       
\end{figure}

\subsection{Other Characteristic Properties of Turbulent Flows}
\label{sec:2.2}%
The model is also consistent with some other characteristic properties of turbulent
flows:

$\bullet$ The dissipation scale $\eta$ obeys the known equation in the Kolmogorov theory
of turbulence
\begin{equation}\label{eq8}
\eta/L \sim {\mathrm {Re}}_{L}^{-3/4}
\end{equation}
Indeed, similarly to $(N/n)^{1/3}$ in Eq. (\ref{eq7a}), the quantity
$q_{i}^{1/3}=(N_{i}/n)^{1/3}$ can be considered to be the Reynolds number of $i$
structure, i.e. $\mathrm{Re}_{i}=q_{i}^{1/3}$. According to the Kolmogorov theory
\cite{Kolmogorov41a}, $\mathrm{Re}_{i} \sim 1$ characterizes the dissipation scale
$\eta$. Let us rewrite $q_{i}=N_{i}/n$ as $(N_{i}/N)\times (N/n)$ and take into account
that $v_{i}/V \sim (l_{i}/L)^{1/3}$ [Eq. (\ref{eq3})], so that $N_{i}/N \sim
[v_{i}l_{i}/(VL)]^{3} \sim (l_{i}/L)^4$. Then,
\begin{equation}\label{eq9}
q_i \sim (N/n)(l_i/L)^4=\mathrm{Re}_{L}^3(l_i/L)^4
\end{equation}
and the equality $q_{i} \sim 1$ (with $l_i \equiv \eta$) leads to Eq. (\ref{eq8}). This,
in particular, suggests that the linear size of the elementary cell could also be
determined as
$\eta v_{\eta}$, where $v_{\eta}$ is the velocity increment at the dissipation scale.%

\begin{figure}[h]
\centering
\resizebox{0.5\columnwidth}{!}{ \includegraphics* {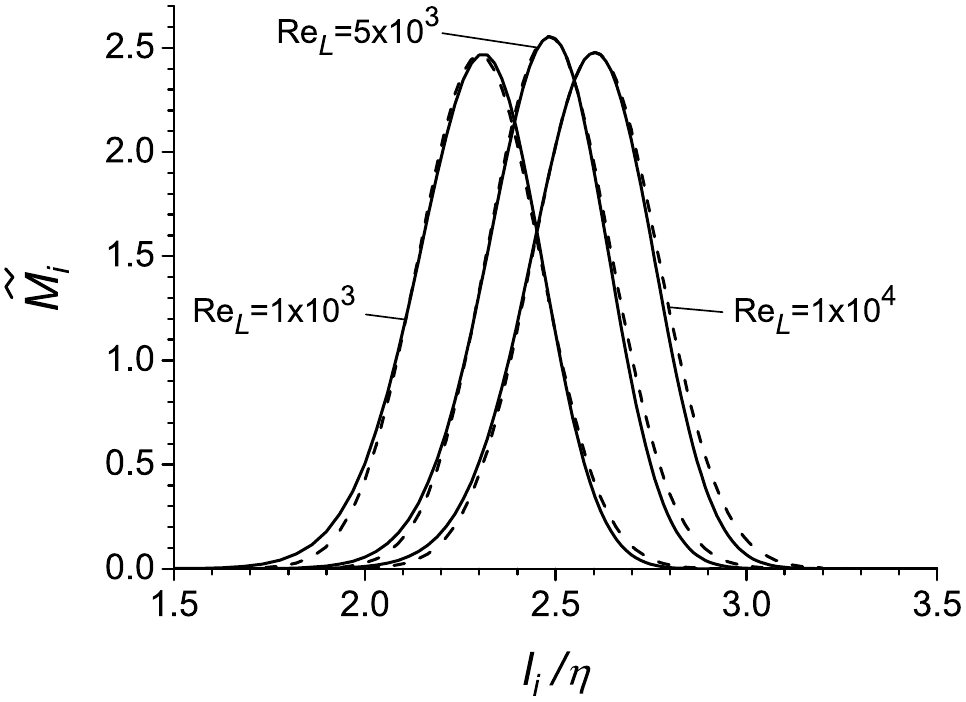}}%
\caption{Characteristic linear size distributions of the localized structures,
$1 \leq q_{i} \leq 200$. Solid lines show the $\tilde{M}_{i}(l_{i}/\eta)$ distributions
averaged over all combinations of $\alpha$ and $\beta$ for which $\mathrm {Re}_{L}$ are
close to those indicated at the curves within 1 percent (typically one of such
distributions dominated). The dashed curves are the
Gaussian fits to the distributions.}
\label{fig4}       
\end{figure}

$\bullet$ The probability density distributions of the linear sizes of the structures
$\tilde{M}_{i}(l_{i}/\eta)$, Fig.\ref{fig4}, are in general agreement with the results of
the direct numerical simulation of isotropic homogeneous turbulence by Jim\'{e}nez {\it
{et al.}} \cite{JimenezWraySaffmanRogallo93}, who found that the vortex radius
distributions are bell-shaped, practically do not shift with $\mathrm {Re}_{L}$ (with the
radius $r$ measured in $\eta$ units), and have maxima at $r \approx 3\eta$. According to
Eqs. (\ref{eq8}) and (\ref{eq9}),  $l_{i}/\eta \approx (q_i)^{1/4}$, so that the maxima
of the distributions at $l_{i}/\eta \approx 2.5$ correspond to $q_i \approx 40$.

$\bullet$ The structure functions and energy spectrum are similar to those in the
Kolmogorov-Obukhov theory \cite{Kolmogorov41a,Obukhov41b}. Since the dependence $v=v(r)$
is assumed to be the same for all localized structures, the structure functions are
independent of the size distributions of localized structures. With Eq. (\ref{eq3}), the
structure function of $p$-th order is $S_{p}(l)=\langle\delta v_{l}^{p}\rangle \sim
l^{p/3}$, where $l$ is the space increment, and the energy spectrum is $E(k) \sim
k^{-5/3}$, where $k=2\pi/l$ is the wave number.

\subsection{Connection with the Hydrodynamic Equations}
\label{sec:2.3}%
The above consideration shows that the proposed model captures characteristic properties
of turbulent flows, in particular, those obtained with the hydrodynamic equations. Then,
the natural question to ask is: Why do two apparently different approaches, the hydrodynamic
equations and the present model, lead to similar results? Or, what is common to these
approaches to lead to similar results? The present model is based on three assumptions:
i) the particles are identical, ii) there exists an elementary cell of the phase space in
which the state of the particles is uncertain, and iii) the kinetic energy of the
fluctuations is as in the Kolmogorov theory for the inertial interval of scales. The
latter assumption seems to be not very essential for our purpose. For example, even if a
quasi-solid rotation of the localized structures (eddies) is supposed, which highly
overestimates their kinetic energy because the absence of the decrease of the velocity
outside the eddy core \cite{Saffman}, the statistical preference of the structured flow
is still preserved (Supplementary Material). The attention should thus be focused on the
former two assumptions. As well known, the Navier-Stokes equation for a gas fluid can be
derived from the Boltzmann equation (the Chapman-Enskog expansion \cite{ChapmanCowling}),
which, in turn, can be obtained from the Liouville equation (see, e.g., Ref.
\cite{FerzigerKaper}). In the transition from the Liouville equation to the Boltzmann
equation, first, the many-particle distribution function is reduced to the
single-particle one under the assumption that the particles are identical (the
Bogoliubov-Born-Green-Kirkwood-Yvon hierarchy, see, e.g. Ref. \cite{Kreuzer}), and,
secondly, because the gas density is taken to be low, the interaction between the
particles is assumed to be a random two-body collision, so that the positions and
velocities of the particles are uncertain within the mean free path $\lambda$ and the
thermal velocity $c$, respectively. It follows that both the assumptions are implicitly
present in the Navier-Stokes equation \cite{liquid}. Therefore, although the
hydrodynamic (Navier-Stokes) equations are incomparably rich in the description of the
turbulent phenomena, because they are able to take into account the specific conditions
under which the transition takes place (the inlet conditions, flow environment, etc.)
and give a dynamic picture of the transition, the driving force for the transition can
have the same statistical origin as in the present model.

\section{Conclusions}
\label{sec:3}%
A model for turbulent fluid flow has been proposed that considers the flow as a
collection of localized spatial structures (eddies) and estimates the statistical weights
of these collections for different Reynolds numbers. The Reynolds number is associated
with the ratio between the total phase volume for the system and that for an elementary
cell in which the state of the particles is uncertain. It has been found
that i) the model successfully explains the onset of turbulence as well as some other
characteristic properties of turbulent flows, and ii) the basic assumptions underlying
the model, i.e. that the particles are identical and an elementary cell of the phase
space exists where the state of the particles is uncertain, are implicitly present in
the Navier-Stokes equation. Considered together, these findings suggest that with all the
variety of dynamic phenomena that the hydrodynamic (Navier-Stokes) equations describe
\cite{Lesieur}, the driving force for the turbulent transition is basically the same as
in the present model, i.e. the system tends to occupy a spatial state that is
statistically dominant at a given Reynolds number. The instability of the flow can then
be a mechanism to initiate the structural rearrangement of the flow to find this state.

The present statistical aspect of the turbulent transition is believed to be important
for a more comprehensive understanding of the nature of this challenging phenomenon as well
as of the other related problems in the field of pattern formation \cite{CrossHohenberg93}.


%
%

\begin{acknowledgements}
I thank B. Ilyushin and D. Sikovsky for useful discussions.
\end{acknowledgements}

\newpage
\begin{center}
{\bf Supplementary Material:}%
\end{center}

\setcounter{figure}{0}
\renewcommand{\thefigure}{S\arabic{figure}}
{\bf Variation of the transition region with parameters $\alpha$ and $\beta$.} Figure
\ref{fig1s} illustrates how the distribution of Fig. 1 at $q_{\mathrm{max}}=200$ changes
with variation of $\alpha$ and $\beta$. The lower boundary shifts to the smaller values
of ${\mathrm {Re}}_{L}$, and the upper boundary to higher values of ${\mathrm {Re}}_{L}$. \\
\begin{figure}[h]
\centering
\resizebox{0.5\columnwidth}{!}{ \includegraphics* {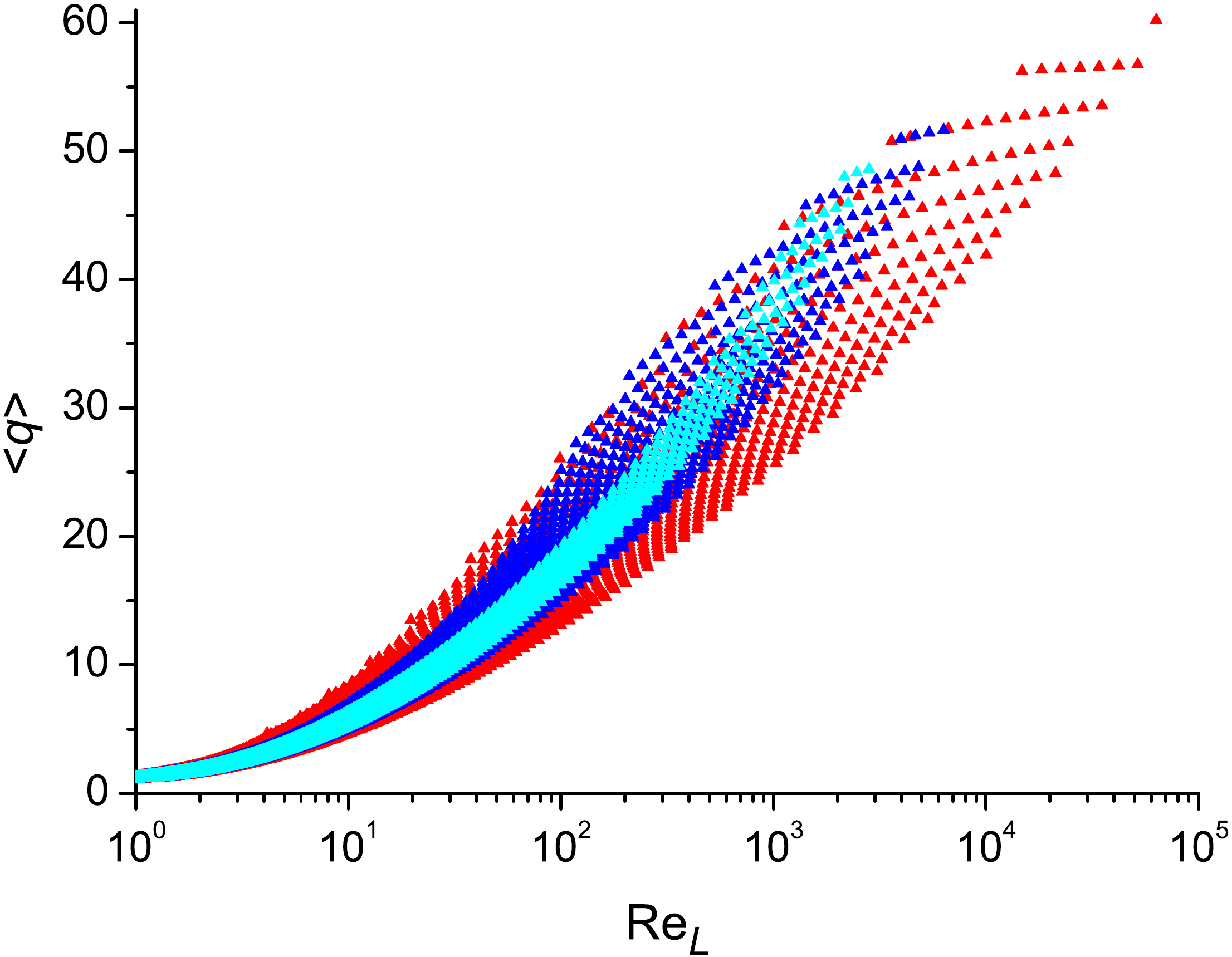}}%
\caption{Average number of the elementary cells in the localized structures $\langle q
\rangle$ as a function of $\mathrm{Re}_{L}$; $1 \leq q_{i} \leq 200$. The cyan triangles
are for $-1.3 \leq \alpha \leq 1.3$ and $-1.3 \leq \beta \leq 1.3$, the blue
triangles are for $-1.5 \leq \alpha \leq 1.5$ and $-1.5 \leq \beta \leq 1.5$ (as in Fig.
1), and the red triangles are for $-2.0 \leq \alpha \leq 2.0$ and $-2.0 \leq \beta \leq
2.0$.}
\label{fig1s}       
\end{figure}

\begin{figure}[h]
\centering
\resizebox{0.49\columnwidth}{!}{ \includegraphics* {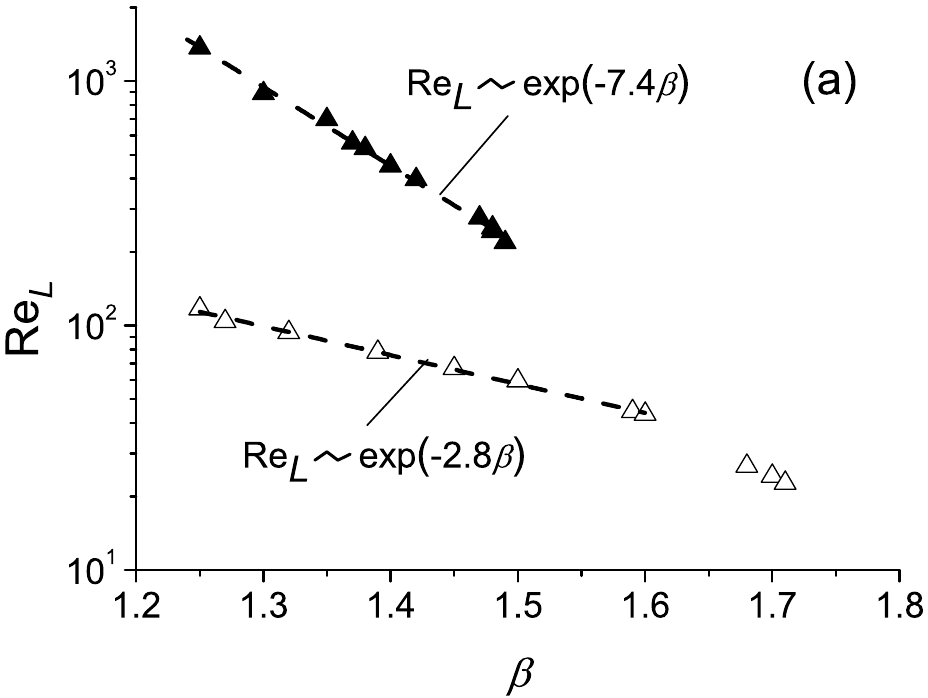}}%
\hfill
\resizebox{0.49\columnwidth}{!}{ \includegraphics* {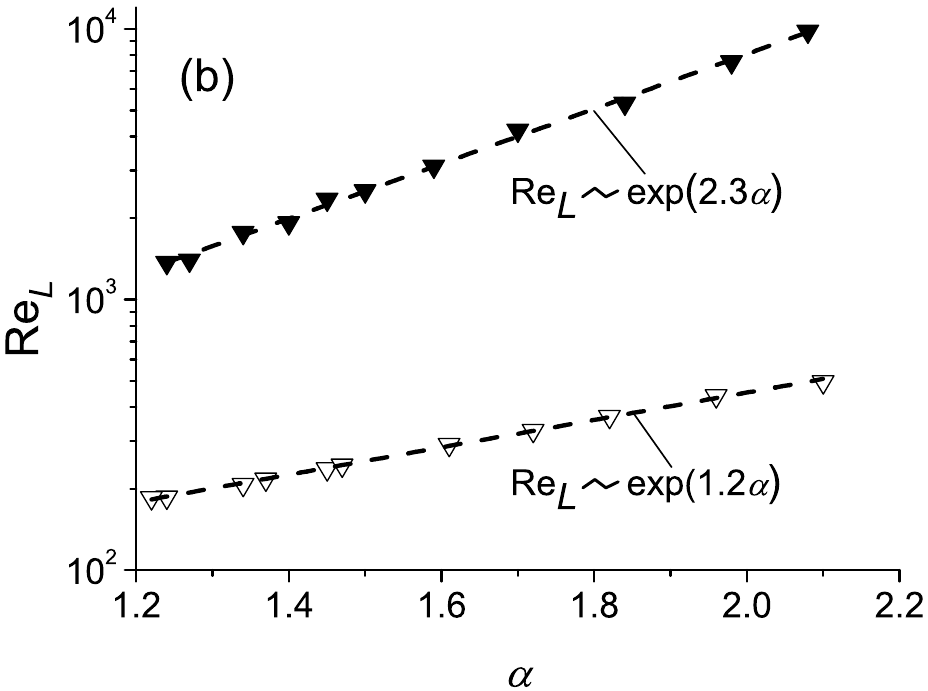}}%
\caption{The Reynolds number ${\mathrm {Re}}_{L}$ corresponding to the lower (panel $\bf
a$) and the upper (panel $\bf b$) boundaries of the transition range as a function of
parameters $\beta$ and $\alpha$, respectively. The empty and solid triangles are for
$\langle q\rangle=20$ and $\langle q\rangle=40$, respectively. The dashed lines show the
best exponential fits to the data.}
\label{fig2s}       
\end{figure}

{\bf Variation of the bounds of the flow stability with the Reynolds number.} Figure
\ref{fig2s} shows how the Reynolds numbers at the lower ({\bf {a}}) and upper
({\bf {b}}) boundaries of the transition range change with $\alpha$ and $\beta$.
While the upper boundary monotonically shifts to larger values of $\mathrm{Re}_L$ as
$\alpha$ increases, the lower boundary shifts to smaller values of $\mathrm{Re}_L$ until
$\beta=\beta^{*}$ is reached ($\beta^{*} \approx 1.725$ at $\langle q\rangle=20$, and
$\beta^{*} \approx 1.5$ at $\langle q\rangle=40$). At larger values of $\beta$ the lower
boundary "freezes", as, e.g., it is seen from Fig. \ref{fig1s}.\\

\begin{figure}
\centering
\resizebox{0.5\columnwidth}{!}{ \includegraphics* {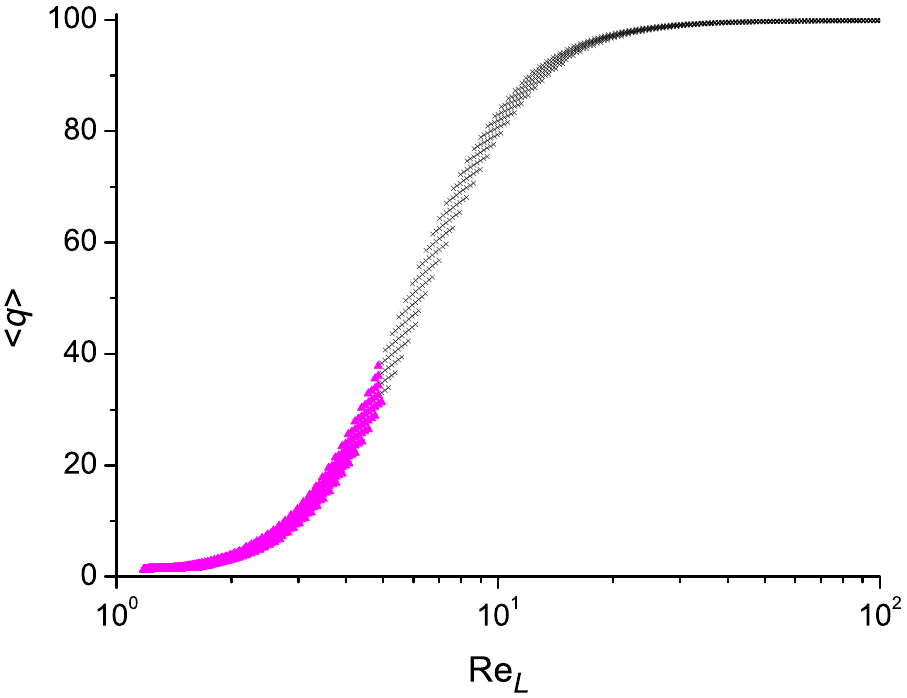}}%
\caption{Average number of the elementary cells in the localized structures $\langle q
\rangle$ as a function of $\mathrm{Re}_{L}$ for quasi-solid rotation of the structures.
$1 \leq q_{i} \leq 100$, and $-0.25 \leq \alpha \leq 0.25$ and $-0.25 \leq \beta \leq
0.25$. Crosses represent all combinations of $\alpha$ and $\beta$, and magenta triangles
stand for the combinations which lead to the physically reasonable bell-shaped
distribution. }
\label{fig3s}       
\end{figure}

{\bf Quasi-solid rotation of the localized structures.} If a quasi-solid rotation of the
localized structures is supposed, in which case $E_{i} \sim N_{i}^{5/3}$, $\langle
q\rangle$ rapidly increases with $\mathrm {Re}_{L}$ as previously, i.e. the statistical
preference of the structured flow is preserved (Fig. \ref{fig3s}). The difference is that the
transition region is located within unreasonably low $\mathrm{Re}$ numbers. This is
because the kinetic energy of the eddies is highly overestimated due to the absence of
the decrease of the velocity outside the eddy core, which is characteristic of more
realistic models, e.g. the Rankine and Lamb-Oseen vortices \cite{Saffman_sm}. \\

\begin{thebibliography}{}
\bibitem{liquid} A similar rigorous derivation of the Navier-Stokes equation for a liquid
fluid is problematic because of the many-body mechanism of interaction between the
particles \cite{Kreuzer}. However, the necessity of the assumptions that the particle are
identical and the system behaves stochastically on the microscale level seems to be
evident.
\end{thebibliography}
\end{document}